\documentclass[fleqn,10pt]{wlscirep}
\usepackage{ifpdf}
\pdfoutput=1
\DeclareRobustCommand{\cbarr}{\includegraphics[height=1.7ex]{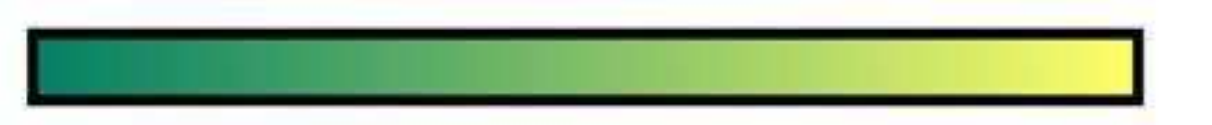}}
\title{Yb- and Er-doped fiber laser Q-switched with an optically uniform, broadband WS$_2$ saturable absorber}
\author[1,*,+]{M. Zhang}
\author[2,+]{G. Hu}
\author[1]{G. Hu}
\author[2]{R. C. T. Howe}
\author[3,*]{L. Chen}
\author[1,4]{Z. Zheng}
\author[2]{T. Hasan}
\affil[1]{School of Electronic and Information Engineering, Beihang University, Beijing, 100191, China}
\affil[2]{Cambridge Graphene Centre, University of Cambridge, Cambridge, CB3 0FA, UK}
\affil[3]{Shenzhen Key Laboratory of Laser Engineering, College of Optoelectronics Engineering, Shenzhen University, Shenzhen, 518060, China}
\affil[4]{Collaborative Innovation Center of Geospatial Technology, Wuhan, 430079, China}
\affil[*]{mengzhang10@buaa.edu.cn}
\affil[*]{l.chen10@szu.edu.cn}
\affil[+]{these authors contributed equally to this work}
\begin{abstract}
We demonstrate a ytterbium (Yb) and an erbium (Er)-doped fiber laser Q-switched by a solution processed, optically uniform, few-layer tungsten disulfide saturable absorber (WS$_2$-SA). 
Nonlinear optical absorption of the WS$_2$-SA in the sub-bandgap region, attributed to the edge-induced states, is characterized by 3.1\% and 4.9\% modulation depths with 1.38 and 3.83 MW/cm$^{2}$ saturation intensities at 1030 and 1558 nm, respectively. 
By integrating the optically uniform WS$_2$-SA in the Yb- and Er-doped laser cavities, we obtain self-starting Q-switched pulses with microsecond duration and kilohertz repetition rates at 1030 and 1558 nm. 
Our work demonstrates broadband sub-bandgap saturable absorption of a single, solution processed WS$_2$-SA, providing new potential efficacy for WS$_2$ in ultrafast photonic applications.
\end{abstract}
\begin{document}
\flushbottom
\maketitle
\thispagestyle{empty}
\section*{Introduction}
Two dimensional (2d) nanomaterials such as mono- or few-layer graphene, semiconducting transition metal dichalcogenides (s-TMDs) and black phosphorous exhibit high third-order optical nonlinear susceptibility and ultrafast carrier dynamics, making them attractive for nonlinear photonics and optoelectronics~\cite{Wang_naturenanotech_2012, bona_natphot_10, Hanlon_arxive_2015}. 
Amongst the 2d materials, s-TMDs are of particular research interest due to their diversity and the distinct yet complementary properties to graphene they offer.
TMDs, a family of $\sim$40 different layered materials, have a general formula MX$_2$, where M is a transition metal atom (e.g. Mo, W or Nb) and X is a chalcogen atom (a group of VI element, e.g. S, Se or Te).
Each TMD layer consists of a single plane of M atoms held between two planes of X atoms by strong covalent bonds.
Depending on the coordination and oxidation states of the M atoms, TMDs may behave as metallic, semiconducting or insulating. 
Similar to other layered materials (LMs), the individual layers in TMD bulk crystals are stacked together by relatively weak van der Waals forces, allowing their exfoliation into single and few layer forms.
The optoelectronic properties of s-TMDs are strongly thickness-dependent. 
For example, the bandgap of s-TMDs typically shifts from indirect for bulk material to direct for monolayer flakes~\cite{Wang_naturenanotech_2012,won_prb_12}. 
Collectively, the bandgaps of s-TMDs span the visible and near-infrared spectrum~\cite{Wang_naturenanotech_2012}.
Such layer-dependent characteristics make s-TMDs comparable or even superior to the zero-gap graphene for a variety of (opto)electronic and photonic applications~\cite{Wang_naturenanotech_2012}.
s-TMDs also offer the possibility of engineering their optical properties for desirable performances~\cite{Hong_naturenanotech_2014}.
We note that while black phosphorous has recently attracted a strong interest~\cite{Hanlon_arxive_2015, Li_arxive_2015, Lu_ol_2015}, poor material stability, even for over a few days, remains a significant drawback~\cite{Hanlon_arxive_2015}.

s-TMDs have been shown to possess remarkable optical and optoelectronic properties, including high optical nonlinear susceptibility~\cite{Kumar_prb_2013, Wang_ami_2013}, ultrafast carrier dynamics~\cite{Wang_acsnano_2013} and broadband working wavelength range~\cite{Zhang_nanores_2015, Woodward_pr_2015}, in addition to robustness and environmental stability.
This has led to the demonstration of numerous nonlinear optical phenomena using s-TMDs, including saturable absorption~\cite{Zhang_nanores_2015, Woodward_pr_2015} (i.e. reduced optical absorption with increased intensity of incident light~\cite{keller_nat_2003}) and optical parametric processes (i.e. second~\cite{ Kumar_prb_2013, janisch_scirep_2014} and third~\cite{Wang_ami_2013} harmonic generation), suggesting that these materials could be a suitable platform for the development of photonic devices.
One such potential application exploiting the saturable absorption property is in the generation of short pulses by mode-locking or Q-switching in laser cavities, where saturable absorber (SA) devices act as a passive optical switch to modulate the intra-cavity loss~\cite{keller_nat_2003}. 
Such short-pulsed or ‘ultrafast’ lasers have become an indispensable tool, playing an increasingly important role in a wide range of applications, including biomedical imaging and therapy, materials processing, fundamental research and military~\cite{Hasan_am_2009}.
Unlike graphene, which has a linear dispersion of Dirac electrons enabling broadband saturable absorption, s-TMDs typically have a (bulk and monolayer) bandgap ranging between $\sim$1-2~eV~\cite{Wang_naturenanotech_2012}.
Some recent progresses have been made in measuring the saturable absorption of s-TMDs in the visible region~\cite{Wang_nanoscale_2014, Zhou_small_2015}, suggesting their potential as a SA device to produce short pulses in this spectral range.

Similar nonlinear optical absorption and fast carrier dynamics in s-TMDs have also been reported in the near-infrared region~\cite{Luo_jlt_2014, Du_scireport_2014, Woodward_arxive_2015, Zhang_nanores_2015}.  
In particular, a number of studies have reported the generation of short pulses from fiber lasers operating at $\sim$ 1.0 ~\cite{Du_scireport_2014,Luo_jlt_2014, Woodward_arxive_2015}, 1.55~\cite{Luo_jlt_2014, Zhang_nanores_2015, Woodward_arxive_2015} and 1.9~$\mu$m ~\cite{Luo_jlt_2014, Woodward_arxive_2015} using several sulfide and selenide-based s-TMDs, including molybdenum disulfide (MoS$_2$) and molybdenum diselenide (MoSe$_2$). 
More recently, tungsten disulfide (WS$_2$) has also been demonstrated to achieve mode-locking and Q-switching, but only in Er-doped fiber lasers~\cite{Mao_scireport_2015, Kassani_ome_2015, Wu_oe_2015, Yan_ome_2015, Khazaeinezhad_cLeo_2015}. 
While the operating wavelengths of these lasers correspond to photon energies below either the bulk or monolayer bandgap of these s-TMDs, the processes behind this sub-bandgap absorption are yet to be fully understood.
Therefore, it is of significant importance to investigate the governing physical mechanism of saturable absorption of WS$_2$ in the near-infrared region and its applicability as a wideband SA material.

Mono- or few-layer WS$_2$ can be produced through a variety of methods, with mechanical cleavage, chemical vapor deposition (CVD) and solution processing techniques such as ultrasonic assisted liquid phase exfoliation (UALPE) being the most commonly exploited ones. 
In particular, UALPE allows mass production of chemically pristine mono- and few-layer WS$_2$ flakes under ambient conditions, without the need for high temperature and complex transfer procedures associated with CVD.
As with the other LMs, WS$_2$ dispersions produced via UALPE can be printed onto optical components such as quartz substrates, mirrors and fiber facets or blended with polymers to form composites for simple integration into a laser cavity, making this strategy very attractive for a wide range photonic and optoelectronic applications~\cite{Hasan_am_2009, Woodward_pr_2015, Sun_physicae_2012}.

Here, we fabricate a few-layer WS$_2$ polymer composite SA based on UALPE for short pulse generation.
The free-standing SA exhibits high spatial uniformity in nonlinear optical properties.
The modulation depth of the WS$_2$-SA is 3.1\% at 1030 nm and 4.9\% at 1558 nm, respectively. 
Using a single WS$_2$-SA composite, we demonstrate microsecond-duration Q-switched pulses with kilohertz repetition rates in all-fiber Yb- and Er-doped lasers to underscore its applicability as a broadband SA material.

\section*{Results}
\subsection*{Sample preparation and characterization}
We fabricate the SA as a freestanding polymer composite film using WS$_2$ nanoflakes produced by UALPE of bulk WS$_2$ crystals. 
The UALPE process for WS$_2$ is similar to that used for other LMs such as graphene~\cite{Sun_acsnano_2010, Hasan_pssb_2010} and several other TMDs~\cite{Woodward_oe_2014, Zhang_nanores_2015, Woodward_pr_2015,Luo_arxive_2015, Hernandez_naturenanotech_2008}, consisting of two steps. 
First, bulk crystals of WS$_2$ are mixed with a suitable solvent and exfoliated via mild ultrasonication. 
This relies on the formation of microbubbles in the solvent resulting from the high-frequency (usually $\sim$20-60kHz) pressure variations~\cite{Mason_appsonochem_2002}. 
Under appropriate frequency, pressure and solvent conditions, these microbubbles become unstable as they grow in size and eventually collapse~\cite{Mason_appsonochem_2002}. 
The shockwaves produced from the collapsed microbubbles create strong shear forces that are sufficient to overcome the weak van der Waals forces between the layers in bulk LM crystals, resulting in exfoliation of thinner flakes. 
The second step involves removal of the unexfoliated, thicker flakes. 
This is typically achieved via centrifugation or filtration~\cite{Hasan_pssb_2010, Coleman_science_2011}.
Thus, UALPE and subsequent processing of WS$_2$ bulk crystals produce exfoliated mono-, bi- and few-layer WS$_2$ flakes in suitable solvents.

The stability of the exfoliated WS$_2$ (and other LMs) dispersion is dependent on minimizing the enthalpy of mixing $\Delta$H~\cite{Coleman_science_2011, Coleman_afm_2009}. 
It has been demonstrated that `good' solvents for exfoliation have dispersive ($\delta_D$), polar ($\delta_P$) and hydrogen bonding ($\delta_H$) Hansen solubility parameters~\cite{hansenbook} that match those empirically derived for WS$_2$ ($\delta_D$$\sim$18~MPa$^{1/2}$, $\delta_P$$\sim$8~MPa$^{1/2}$, $\delta_H$$\sim$7.5~MPa$^{1/2}$)~\cite{Cunningham_acsnano_2012}. 
However, the best-suited solvents (e.g. N-cyclohexyl-2-pyrrolidone [CHP], cyclohexanone and N-methyl-2-pyrrolidone [NMP]) are challenging for device/composite processing due to their relatively high boiling points (NMP~$\sim$202$^{\circ}$C, cyclohexanone~$\sim$155$^{\circ}$C, CHP$\sim$154$^{\circ}$C). 
UALPE in pure lower boiling point solvents such as alcohols and water does not typically yield stable dispersions due to relatively strong mismatch between $\delta_D$, $\delta_P$ and $\delta_H$ of these solvents to those of WS$_2$. 
In this case, surfactants can be used to stabilize the dispersion. 
Such stable dispersions are necessary for the fabrication of optically homogeneous composite SAs, as the slow drying process would otherwise allow aggregates to form~\cite{Hasan_am_2009}, leading to scattering losses and unreliable device performance~\cite{Wong_apl_2004, Hasan_am_2009}. 
This approach has therefore been widely used to fabricate 1d and 2d material-based SAs~\cite{Sun_acsnano_2010,Hasan_pssb_2010, Woodward_arxive_2015, Sun_physicae_2012, Martinez_np_2013}.

It has been experimentally shown with graphene that quasi-2d surfactants with hydrophobic and hydrophilic faces (e.g. sodium deoxycholate [SDC], a bile salt surfactant) are well-suited to exfoliating and stabilizing hydrophobic 2d materials (such as WS$_2$) in water~\cite{Hasan_pssb_2010}. 
For stable dispersions, the surfactant should be present at a concentration in excess of the temperature dependent critical micelle concentration (CMC), defined as the surfactant concentration above which its molecules can spontaneously self-arrange to form micelles in water~\cite{Butt_2006}. 
Therefore, if exfoliated hydrophobic flakes (such as WS$_2$) are present in a water-surfactant solution above the corresponding CMC value, the surfactant molecules are expected to encapsulate and stabilize the flakes, preventing their reaggregation and sedimentation~\cite{Coleman_afm_2009,Hasan_pssb_2010,Lotya_joacs_2009, Smith_am_2011}.

We prepare the stable few-layer WS$_2$ dispersion by mixing 100 mg WS$_2$ crystals with $\sim$70 mg SDC surfactant in 10 mL of DI water and sonicating for 12 hours in a bath sonicator at $\sim$15 $^{\circ}$C. 
We select SDC, a di-hydroxy bile salt surfactant over the more commonly used sodium cholate (SC), a tri-hydroxy bile salt due to its higher hydrophobic index, which should allow stronger interaction between the surfactant and WS$_2$\cite{Hasan_pssb_2010}.
The CMC of SDC in water at room temperature is found to be 4.7 mM ($\sim$2~g.L$^{-1}$) by pendant droplet measurement of the surface tension of different concentration SDC solutions (the surface tension of solutions changes rapidly below the CMC, but stabilizes above).
Thus the surfactant concentration we use here is $\sim$3.5 times the CMC, sufficient to adequately support the exfoliated WS$_2$~\cite{Hasan_pssb_2010, Sugihara_langmuir_2000,Miyajima_jcs_1988}. 
The sonicated dispersion, containing a mixture of exfoliated and unexfoliated materials, is centrifuged at $\sim$1500~$g$ for one hour. The top 70\% of the dispersion, enriched with mono-, bi- and few layer flakes, is then decanted for characterization and SA fabrication.

\subsubsection*{Dispersion characterization}
\begin{figure}[ht!]
	\centering
	\includegraphics[width=0.73\textwidth]{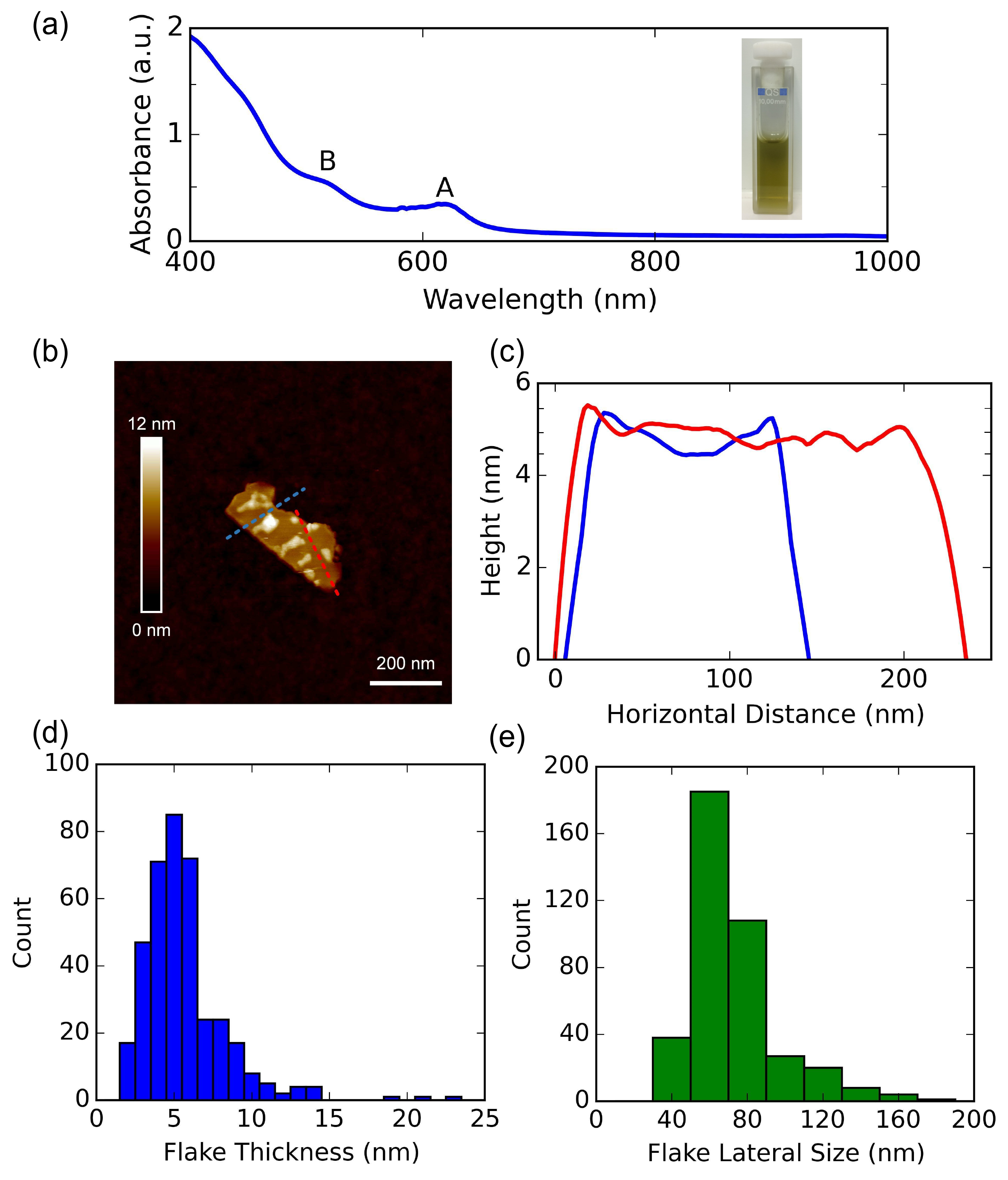}
	\caption{(a) Linear optical absorption of WS$_2$ flakes in a 10\% v/v dispersion. Inset, Stable WS$_2$ dispersion (diluted to 10\% v/v) in a cuvette: (b) AFM image of a typical WS$_2$ flake; (c) height variations of the flakes along the marked line; (d) TEM distribution of flake thickness; (e) distribution of lateral WS$_2$ flake dimensions.}
	\label{dispersion}
\end{figure}

The optical absorption spectrum of the WS$_2$ dispersion, diluted to 10\% v/v to reduce scattering effects~\cite{Bohren_2008}, is shown in Fig.~\ref{dispersion}(a), with the inset of a photograph of the cuvette containing the dispersion.
The spectrum shows the characteristic WS$_2$ excitonic peaks, referred to as A (at $\sim$630~nm) and B (at $\sim$520~nm) according to common nomenclature~\cite{Beal_jopc_1972}. 
The spectrum can be used to estimate the concentration of WS$_2$ using Beer-Lambert law ($A_{\lambda}= \alpha_{\lambda} cl$), where $c$ is the WS$_2$ concentration (gL$^{-1}$), $l$ is the distance the light passes through the dispersion (m) and $A_{\lambda}$ and $\alpha_{\lambda}$ are the absorption (a.u.) and material dependent optical absorption coefficient (Lg$^{-1}$m$^{-1}$) at wavelength $\lambda$ (nm), respectively. 
Using a combination of optical absorption spectroscopy and thermogravimetric analysis on a set of dispersions~\cite{Zhang_nanores_2015}, we estimate the value of $\alpha_{630}$ for WS$_2$ $\sim$1324 Lg$^{-1}$m$^{-1}$.
From this, we estimate the concentration of dispersed WS$_2$ to be 0.23 gL$^{-1}$.

The distribution of flake thicknesses and lateral dimensions is measured via atomic force microscopy (AFM).
Samples are prepared by drop-casting a diluted (5\% v/v) WS$_2$ dispersion onto a Si/SiO$_2$ wafer. 
The wafers are then rinsed with DI water to remove residual surfactant, giving clean and isolated flakes on the substrates. 
A typical flake is shown in Fig.~\ref{dispersion}(b). 
The corresponding height profile, presented in Fig.~\ref{dispersion}(c), shows a thickness of $\sim$5 nm.
The average thickness [Fig.~\ref{dispersion}(d)], measured across $\sim$400 individual flakes, is (5.0$\pm$0.1) nm.
We find $\sim$73\% of the flakes are $\leq$5 nm thick, corresponding to $<$8 layers, assuming $\sim$1 nm for a monolayer flake and 0.6 nm for each subsequent layer~\cite{Gutierrez_nanolett_2012}. 
As shown in Fig.~\ref{dispersion}(e), the average lateral dimension of the flakes is (62$\pm$1) nm.

\subsubsection*{Composite characterization}
\begin{figure}[ht!]
	\centering
	\includegraphics[width=\linewidth]{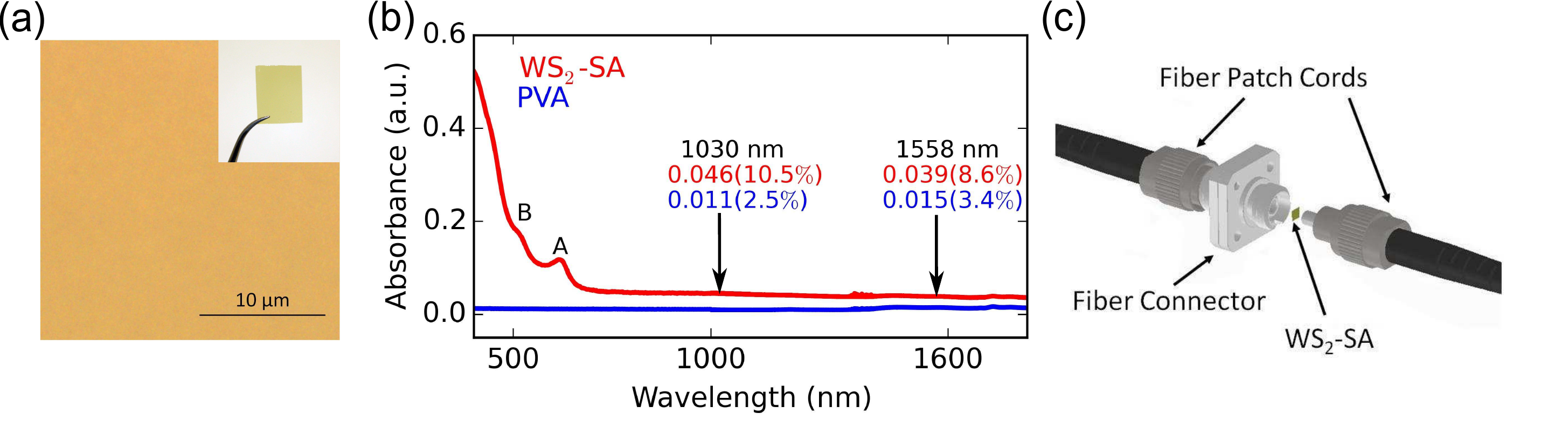}
	\caption{(a) Optical micrograpgh of the free-standing WS$_2$-PVA SA composite, confirming absence of aggregates. Inset, photograph of WS$_2$-SA. (b) Optical absorbance of a pure PVA and WS$_2$-PVA SA. (c) Schematic showing integration of WS$_2$-PVA SA device between two fiber patch cords.}
	\label{composite}
\end{figure}

To allow integration of the SA into the fiber laser cavity, a free-standing composite film is prepared from the WS$_2$ dispersion using an approach previously demonstrated for other 1d and 2d materials~\cite{Hasan_am_2009, Sun_acsnano_2010,Hasan_pssb_2010, Woodward_oe_2014, Woodward_pr_2015, Woodward_arxive_2015, Zhang_nanores_2015, Sun_physicae_2012, Martinez_np_2013}. 
The composite is prepared by homogeneously mixing the WS$_2$ dispersion with a 5 wt\% aqueous solution of polyvinyl alcohol (PVA) polymer. 
PVA is used as the host polymer because it does not exhibit strong optical absorption at 1030 and 1558 nm, its solvent compatibility (PVA is water soluble), ease of processability (can be processed at room temperature), robustness and flexibility (dried thin films have high tensile strength and are not brittle). 
The mixture is poured into a Petri dish and allowed to dry at room temperature in a desiccator, producing a $\sim$30~$\mu$m thick, free-standing WS$_2$-polymer composite SA [Fig.~\ref{composite}(a) inset].

For transmissive type thin film SA devices, optical uniformity without defects (such as cracks, voids, material aggregates etc.) is an important consideration to ensure reliable and repeatable device performance.
The uniformity and quality of the WS$_2$-SA is assessed by using optical microscopy, a commonly used technique employed for other nanomaterial based SAs~\cite{Hasan_am_2009, Woodward_arxive_2015}. 
The optical micrograph, shown in Fig.~\ref{composite}(a), confirms the absence of $>$1 $\mu$m aggregates in the SA composite, which would otherwise have led to undesirable nonsaturable scattering losses~\cite{Bohren_2008}.

Figure~\ref{composite}(b) shows the optical absorption of the composite film. 
Measurement of a pure PVA film prepared by the same method is also included in Fig.~\ref{composite}(b) as a reference. 
We note that the WS$_2$-SA composite shows non-zero absorption, even at energies below bandgap of either bulk ($\sim$1.3-1.4 eV, 954-886 nm~\cite{Wang_naturenanotech_2012, Kam_jopc_1982, Kuc_prb_2011}) or monolayer ($\sim$2.1 eV, 590 nm~\cite{Kuc_prb_2011, Wang_naturenanotech_2012}) WS$_2$. 
Considering contributions from PVA (0.011-0.015 absorbance, 2.5-3.4\%) [Fig.~\ref{composite}(b)] the WS$_2$-PVA composite shows a marked increase in absorption (0.046-0.039 absorbance, 10.05\%-8.6\%), indicating strong optical absorption from the embedded WS$_2$ flakes at both 1030 (1.2 eV) and 1558 nm. 
We have recently attributed this sub-bandgap light absorption in other s-TMD materials (MoS$_2$~\cite{Woodward_pr_2015, Zhang_nanores_2015,Roxlo_jovsta_1987}, MoSe$_2$~\cite{Woodward_arxive_2015}) to edge-induced sub-bandgap states~\cite{Woodward_oe_2014, Zhang_nanores_2015, Woodward_pr_2015, Woodward_arxive_2015}. 
We propose that similar edge-induced states arise within the material bandgap in our UALPE WS$_2$ flakes, leading to the observed sub-bandgap light absorption.

\begin{figure}[ht!]
	\centering
	\includegraphics[width=0.91\textwidth]{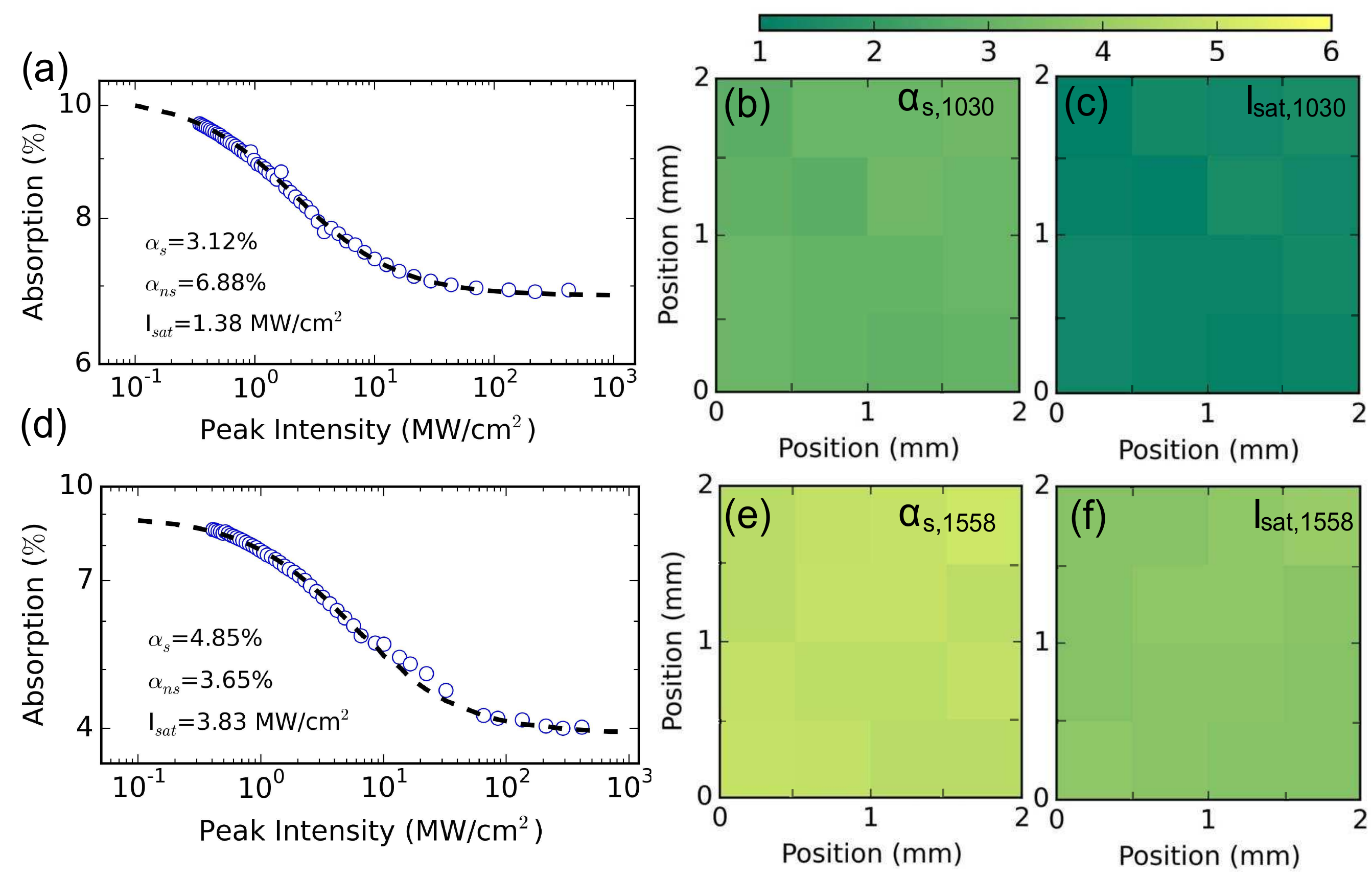}
	\caption{Nonlinear optical absorption at (a)-(c) 1030 nm and (d)-(f) 1558 nm measured by an open aperture Z-scan technique.  (a), (d) Typical datasets from Z-scan experiment. Mapping of the homogeneity of the sample used in the experiment: variation in modulation depth (b), (e),  colorbar:~1~\cbarr~$6\mathrm{\%}$;  and saturation intensity (c), (f), colorbar:~1~\cbarr~$6~\mathrm{MW/cm^{2}}$.}
	\label{zscan}
\end{figure}

The nonlinear optical absorption of the WS$_2$-SA is characterized using an open-aperture Z-scan technique. 
For this, an ultrashort fiber source operating at 1030 nm is used as the pump light (120 fs pulse duration, 20 MHz pulse repetition rate), split using a 90\%:10\% fused fiber coupler, the latter enabling monitoring of power used as a reference. 
The SA composite is swept through the focus of a beam of the remaining port and the transmitted power is recorded as a function of incident intensity on the device. 
A typical dataset from a single Z-scan measurement, at a fixed transverse position on the WS$_2$-SA, shown in Fig.~\ref{zscan}(a), can be well-fitted with the two-level SA model~\cite{Haus75}. 
From the fit, the following SA parameters are extracted: at 1030 nm, the saturation intensity, I$_{sat,1030}$$\sim$1.38 MW/cm$^{2}$, modulation depth, $\alpha_{s,1030}$$\sim$3.1\% and nonsaturable absorption, $\alpha_{ns,1030}\sim$6.9\%. 
The same measurement is also carried out at the wavelength of 1558 nm (150 fs pulse duration, 10 MHz pulse repetition rate). 
The measured parameters are: I$_{sat,1558}$ 3.83 MW/cm$^{2}$, $\alpha_{s,1558}$ $\sim$4.9\%, and $\alpha_{ns,1558}$ $\sim$3.7\%, respectively [Fig~\ref{zscan}(d)]. 
Thus, the WS$_2$-SA shows strong saturable absorption in both these wavelengths.

The uniformity of the WS$_2$ flakes embedded in the free-standing WS$_2$-PVA composite is next evaluated by raster scanning the sample.
This is essential to ensure reliable and repeatable performances of the WS$_2$-SA devices.
For this, we measure the nonlinear saturation of the $\sim$30~$\mu$m SA with 0.5 mm spatial increments in the XY plane, across a 2$\times$2~mm section, at both wavelengths. 
The extracted data, processed to display the variation in $\alpha_{s,1030}$ and $\alpha_{s,1558}$ [Fig.~\ref{zscan}(b) and (e)] and I$_{sat,1030}$, I$_{sat,1558}$ [Fig.~\ref{zscan}(c) and (f)] are presented on two-dimensional grids. 
The spatial increment, i.e. the resolution in the scanning, is limited by the laser spot size during the measurement. Figure~\ref{zscan}(b), (c), (e), (f) show clear evidence of optical homogeneity (with standard deviations of 0.13\%, 0.17 MW/cm$^{2}$, 0.13\%, 0.11 MW/cm$^{2}$, respectively), allowing us to obtain repeatable performance across the WS$_2$-SA sample.

\subsection*{Demonstration of Q-switching a fiber laser using few-layer WS$_2$-SA}
The demonstrated sub-bandgap saturable absorption of the few-layer WS$_2$-PVA composite at 1030 and 1558 nm indicates that the device could be used to modulate the loss and Q-factor of a fiber laser cavity. 
This could in turn to be exploited to generate a regular train of Q-switched pulses in this spectral region.
To explore the potential of using a single WS$_2$-SA for short-pulse generation at different wavelengths, fully fiber-integrated Yb- and Er-doped lasers are constructed. 
For each laser cavity, a ring configuration is adopted, consisting of entirely isotropic, single-mode fiber. 
The Yb and Er fiber amplifiers consist of single-mode Yb- and Er-doped active fiber, respectively, co-pumped by a 974 nm pump diode. 
In addition to the fiber amplifier, each cavity includes a polarization-independent optical isolator to ensure unidirectional propagation, fused-fiber output coupler for both spectral and temporal diagnostics and polarization controller to adjust the net cavity birefringence [Fig.~\ref{Yb_laser}(a) and ~\ref{Er_laser}(a)]. 
The WS$_2$-SA is integrated into the cavities by sandwiching a $\sim$1~mm$\times$1~mm of the composite between two fiber patch chords, shown in Fig.~\ref{composite}(c). 
The total cavity length for the Yb- and Er-doped lasers is 66 m and 9 m, respectively.

\subsubsection*{Q-switched Yb-doped fiber laser characterization}
\begin{figure}[ht!]
	\centering
	\includegraphics[width=0.86\textwidth]{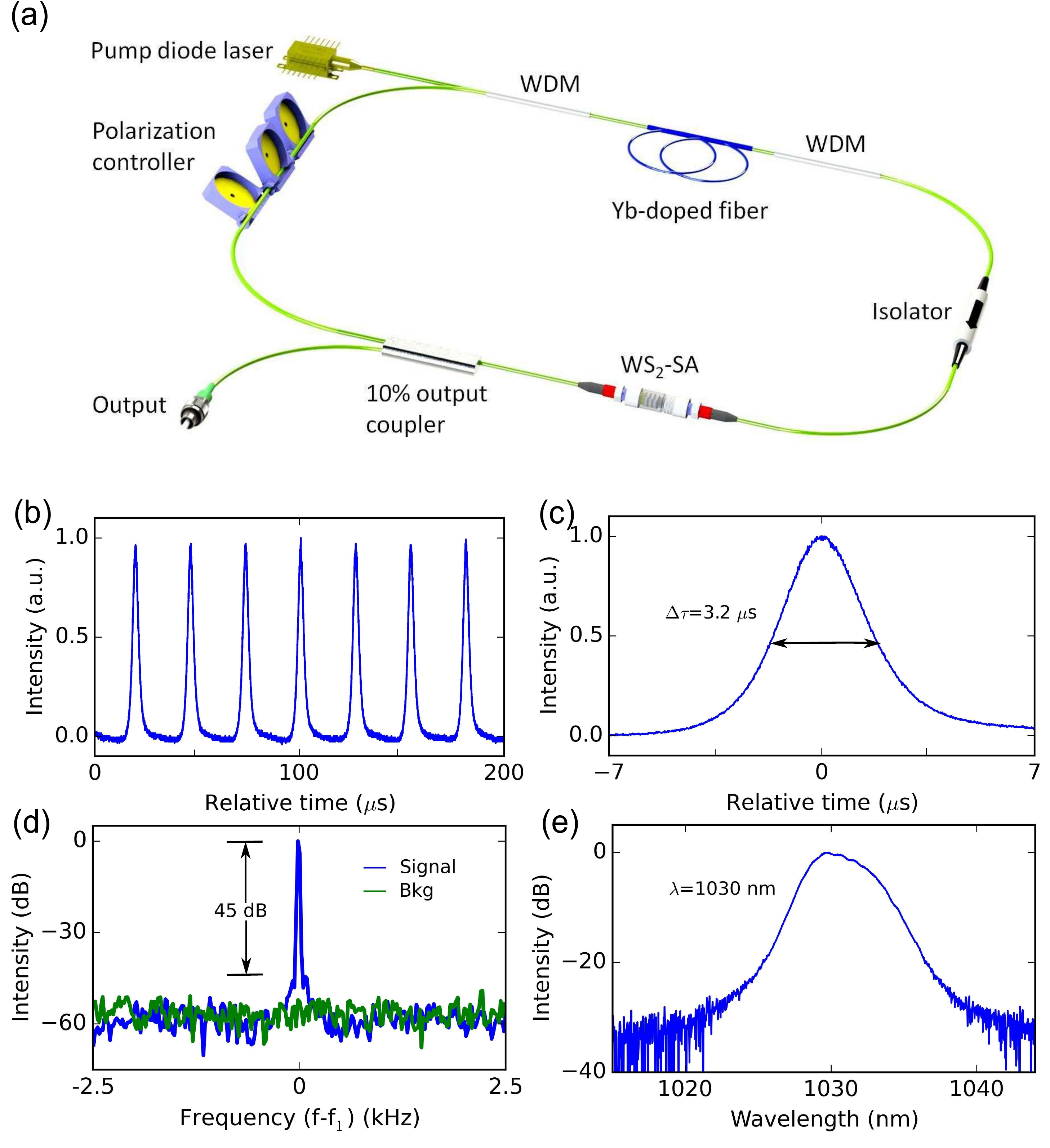}
	\caption{Q-switched Yb-doped fiber laser: (a) cavity configuration; (b) output pulse train, with a spacing of 27.2 $\mu$s; (c) single pulse profile, with 3.2 $\mu$s FWHM pulse width; (d) radio frequency spectrum of fundamental frequency on a 5 kHz span, where $\mathrm{f_1=36.7 kHz}$, with the green trace showing the noise floor of the RF analyzer; (e) measured optical spectrum.}
	\label{Yb_laser}
\end{figure}

Self-starting Q-switching is obtained from the all-fiber integrated Yb-doped laser, generating a stable train of pulses, centered at 1030 nm [Fig.~\ref{Yb_laser}(e)]. 
Typical output characteristics of the laser, at 0.5 mW average output power, are shown in Fig.~\ref{Yb_laser}. 
Pulses are generated with 27.2~$\mu$s spacing, corresponding to 36.7 kHz repetition rate [Fig.~\ref{Yb_laser}(b)] and a full width at half maximum (FWHM) pulse duration of 3.2~$\mu$s [Fig.~\ref{Yb_laser}(c)]. 
The radio frequency (RF) spectrum of the output shows a high signal-to-background contrast of 45 dB [Fig.~\ref{Yb_laser}(d)], indicating good pulse train stability, comparable to Q-switched fiber lasers based on other 2d layered materials~\cite{Liu_ol_2011, Popa_apl_2011, Woodward_pr_2015}.

The pulse properties in continuous-wave pumped Q-switched lasers rely on nonlinear dynamics in the gain medium and SA. 
This leads to a dependence of cavity repetition rate and pulse duration on pump power~\cite{Degnan_jqe_1995}. 
A pulse is emitted once the storage energy of the cavity reaches a certain threshold.
Therefore, a greater pump power enables higher repetition rates and  results in shorter pulses. 
This is experimentally observed by changing the pump power as the pulse duration is reduced from 6.4 $\mu$s to 3.2 $\mu$s and the cavity repetition rate is increased from 24.9~kHz to 36.7~kHz [Fig.~\ref{variation}(a)]. 
The maximum pulse energy is 13.6~nJ, limited by the available pump power. 
We believe that higher pulse energies could be achieved by further optimizing the laser cavity.
The pulse duration could also be further shortened by reducing the length of the laser cavity.

\subsubsection*{Q-switched Er-doped fiber laser characterization}
\begin{figure}[h!]
	\centering
	\includegraphics[width=0.86\textwidth]{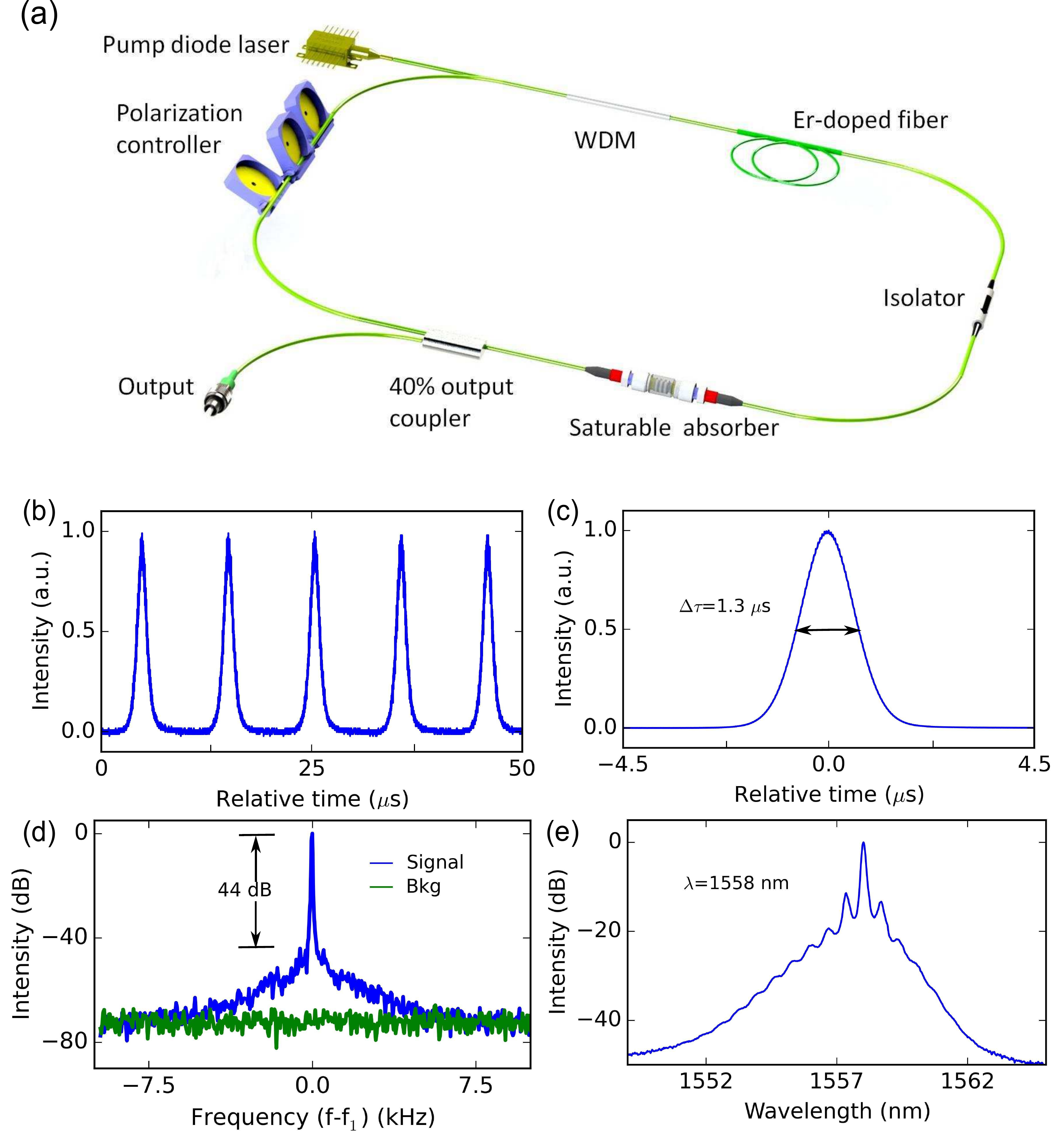}
	\caption{Q-switched Er-doped fiber laser: (a) cavity configuration; (b) output pulse train, with a spacing of 10.3~$\mu$s; (c) single pulse profile, with 1.3~$\mu$s FWHM pulse width; (d) radio frequency spectrum of fundamental frequency on a 20 kHz span, where $\mathrm{f_1=97 kHz}$, with the green trace showing the noise floor of the RF analyzer; (e) measured optical spectrum.}
	\label{Er_laser}
\end{figure}

\begin{figure}[h!]
	\centering
	\includegraphics[width=\textwidth]{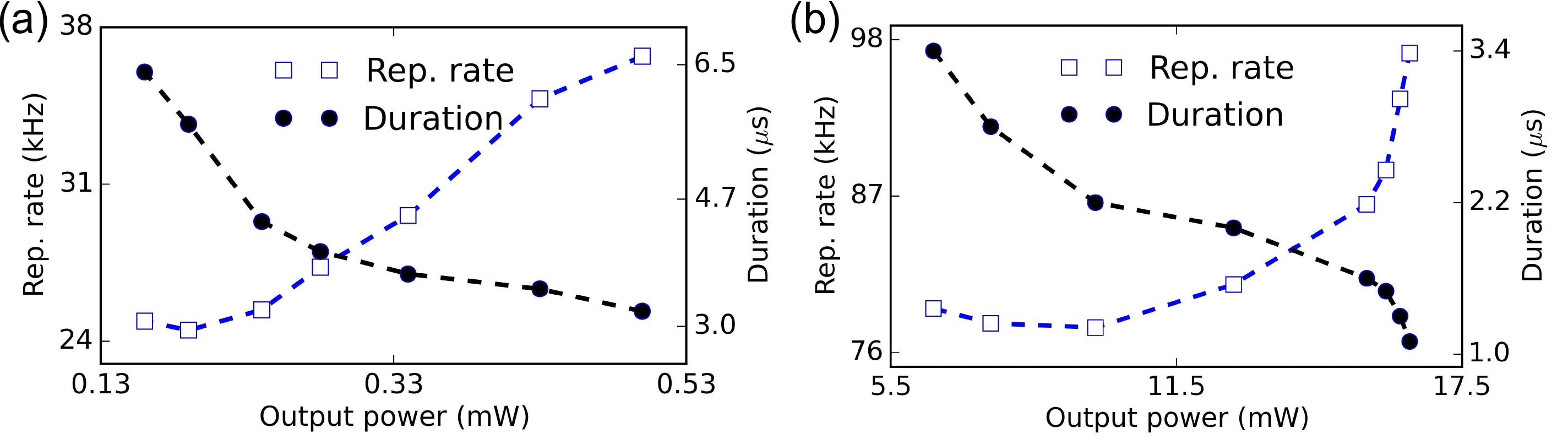}
	\caption{Variation of the pulse duration and repetition rate with average output power for Q-switched (a) Yb-, (b) Er-doped fiber lasers using few-layer WS$_2$-PVA composite SA.}
	\label{variation}
\end{figure}

To demonstrate the sub-bandgap saturable absorption for short pulse generation at a longer wavelength, the same SA is subsequently incorporated into the Er-doped fiber laser described above. 
Self-starting Q-switching operation is achieved at 1558 nm [Fig.~\ref{Er_laser}(e)] with an average output power of 6.4~mW. 
Typical output pulse train properties are shown in Fig.~\ref{Er_laser}(b). 
The cavity repetition rate is 97.1 kHz, corresponding to a pulse-to-pulse spacing of 10.3~$\mu$s.
Figure~\ref{Er_laser}(c) shows the single pulse profile, with a FWHM pulse duration of 1.3 $\mu$s. 
The signal-to-noise background contrast of RF spectrum, plotted in Fig.~\ref{Er_laser}(d) is 44 dB, again, showing the good pulse train stability of the Er-laser. 
With increasing pump power, the repetition rate is tuned from 79 kHz to 97 kHz, with corresponding pulse duration from 3.4~$\mu$s to 1.1~$\mu$s as the average output power is increased from 6.4 mW to 16.4 mW [Fig.~\ref{variation}(b)]. 
The pulse energy is 179.6~nJ at the highest output power.

\section*{Discussion}
Following stable Q-switching results obtained by using the WS$_2$-SA composite at 1030 and 1558 nm, the same experiment is conducted with a $\sim$30~$\mu$m-thick pure PVA film (fabrication process similar to that used for WS$_2$-SA fabrication, but without the WS$_2$ flakes). 
No Q-switching is observed at any power level or polarization controller position, confirming that the sub-bandgap saturable absorption arises from the few-layer WS$_2$ flakes.

As discussed before, the WS$_2$-SA exhibits non-zero sub-bandgap absorption at energies below the bulk ($\sim$1.3-1.4 eV, 954-886 nm~\cite{Wang_naturenanotech_2012, Kuc_prb_2011, Kam_jopc_1982}) or monolayer ($\sim$2.1 eV, 590 nm~\cite{Wang_naturenanotech_2012, Kuc_prb_2011}) WS$_2$ bandgap. 
The sub-bandgap absorption and its nonlinear behavior is also verified by Z-scan measurement and demonstration of Q-switched lasers at 1030 (1.2 eV) and 1558 nm (0.8 eV).
Similar observation of sub-bandgap saturable absorption in other s-TMDs have also been reported~\cite{Du_scireport_2014, Luo_jlt_2014, Zhang_nanores_2015, Woodward_arxive_2015}.
For the case of MoS$_2$ and MoSe$_2$, we recently proposed that the sub-bandgap absorption arises from edge-induced sub-bandgap states~\cite{Woodward_pr_2015, Zhang_nanores_2015, Woodward_arxive_2015}. 
Our proposal is based on previous experimental observations of increased sub-bandgap absorption in lithographically patterned MoS$_2$ compared to their large crystals due to increased edge to surface area ratio~\cite{Roxlo_jovsta_1987}.
From the AFM data and considering approximately square flakes for ease of calculation, we estimate a high 1:6 ratio between the edge ({\it = 4$\times $d$\times$ t} where {\it d} is the lateral dimension and {\it t} is the thickness of the flakes) and surface area ({\it =2$\times d^2$}) of the WS$_2$ flakes. 
We thus suggest that the observed sub-bandgap absorption in the WS$_2$-SA is also due to the edge-states, promoted by the high edge to surface area ratio of the UALPE WS$_2$ flakes.
This edge-induced sub-bandgap absorption can be saturated at high incident intensities by Pauli blocking, which enables WS$_2$ to act as an SA material in the near-infrared region. 
Additionally, a distribution of edge-induced states within the bandgap could explain the wideband saturable absorption experimentally observed here, and in recent reports at a number of different laser wavelengths in other s-TMDs~\cite{Du_scireport_2014, Luo_jlt_2014, Zhang_nanores_2015, Woodward_arxive_2015, Woodward_pr_2015}.

We note that Wang et al. have also proposed a complementary explanation for this phenomenon based on crystallographic defect states, supported by theoretical bandgap studies by varying the ratio of M and S atoms~\cite{Wang_am_2014}.
Zhang et al.~\cite{Zhang_acsnano_2015} reported nonlinear optical measurements of WS$_2$ and MoS$_2$, grown by high temperature, direct vapor phase sulfurization of pre-deposited metal films.
Both materials showed layer dependent saturable and reverse saturable absorption (RSA, due to two-photon absorption) properties.
Additionally, the authors demonstrated RSA for 1-3, 18-20, 39-41 layered samples at 1030 nm while SA for 18-20 layers and RSA for 1-3 layer WS$_2$ at 800 nm.
We do not observe any such RSA behavior in our samples ($<$8 layers) during the nonlinear measurements (maximum peak intensity, 420 MW/cm$^2$) and Q-switching operation.
Another recent report attributes exciton-exciton interaction in bulk and monolayer MoSe$_2$ (a selenium based s-TMD) for saturable absorption at excitonic resonance~\cite{Kumar_prb_2014}.
However, this does not explain the saturable sub-bandgap absorption we observe in our experiments here.
A distribution of edge-induced states within the bandgap could better explain the wideband saturable absorption, supported by recent reports on mode-locking or Q-switching at a number of different laser wavelengths in other s-TMDs~\cite{Woodward_arxive_2015, Woodward_pr_2015}. 
We note that grain boundaries and other defects present in s-TMD crystals, such as those grown by CVD may also strongly contribute to such sub-bandgap absorption.
Indeed, this may explain ultrafast pulse generation using CVD-grown s-TMDs below the bulk and monolayer bandgap~\cite{Brien_scireport_2014, Zhang_nanolett_2014, Lv_acr_2014, Liu_nanolett-2012}.
Considering the above discussion, we propose that the large edge-to-surface ratio of nanoflakes of WS$_2$, in particular, prepared by UALPE, would be the primary origin of the optical absorption below the fundamental material bandgap.
In summary, a free-standing few-layer WS$_2$-PVA SA has been fabricated by UALPE of chemically pristine WS$_2$. 
Using this SA, we have developed self-starting Q-switched Yb- and Er-doped fiber lasers for short pulse generation. 
We have proposed edge-induced sub-bandgap states in WS$_2$ as the primary reason for broadband saturable absorption in the near IR spectral region. 
This extends our existing understanding for this phenomenon to a wider class of s-TMDs with regard to their potential for future photonic technologies.

\section*{Acknowledgements}
The authors thank E. J. R. Kelleher for valuable discussions. MZ acknowledges support from Beihang University, China, through a Zhuoyue Bairen Program and TH from the Royal Academy of Engineering through a fellowship (Graphlex). 
\section*{Author contributions statement}
M. Z., L. C. and T. H. conceived the experiments, G. H., G. H., R. C. T. H. conducted the experiments, M. Z., G. H., R. C. T. H. and T. H. analyzed the results, M. Z., L. C., Z. Z. and T. H. wrote the manuscript. All authors reviewed the manuscript.
\section*{Additional information}
\textbf{Competing financial interests.}{~The authors declare no competing financial interests.} 
\end{document}